\documentclass[twocolumn,times]{aastex63}
\turnoffediting

\newcommand{\msun}{\ensuremath{M_\sun}}
\received{14 June 2022}
\revised{15 August 2022}
\accepted{DD MMMM YYYY}
\submitjournal{ApJ}
\shorttitle{Resolution study}
\shortauthors{Rivas et al.}
\let\tablenum\relax
\usepackage{siunitx}
\usepackage[version=4]{mhchem}
\usepackage{gensymb}
\DeclareSIUnit\erg{erg}
\begin{document}
\title{The Impact of Resolution on Double-Detonation Models for Type Ia Supernovae}
\newcommand{\UTphys}{Department of Physics and Astronomy, University of Tennessee, Knoxville, TN 37996-1200, USA}
\newcommand{\NCCS}{National Center for Computational Sciences, Oak Ridge National Laboratory, P.O. Box 2008, Oak Ridge, TN 37831-6164, USA}
\newcommand{\ORNLphys}{Physics Division, Oak Ridge National Laboratory, P.O. Box 2008, Oak Ridge, TN 37831-6354, USA}
\correspondingauthor{Fernando Rivas}
\email{frivas1@vols.utk.edu}
\author[0000-0002-6762-6070]{Fernando Rivas}
\affiliation{\UTphys}
	\author[0000-0003-3023-7140]{J. Austin Harris}
	\affiliation{\NCCS}
\author[0000-0002-9481-9126]{W. Raphael Hix}
	\affiliation{\ORNLphys}
	\affiliation{\UTphys}
\author[0000-0002-5358-5415]{O. E. Bronson Messer}
	\affiliation{\NCCS}
	\affiliation{\ORNLphys}
	\affiliation{\UTphys}
\begin{abstract}

Thermonuclear supernovae are the result of the violent unbinding of a white dwarf, 
but the precise nature of the explosion mechanism(s) is a matter of active debate.
To this end, several specific scenarios have been proposed to explain the observable traits of SNe Ia. 
A promising pathway is the double-detonation scenario, where a white dwarf accretes a shell of helium-rich material from a companion and a detonation in the resulting helium shell is the primary cause of the explosion. 
Through a set of two-dimensional grid-based simulations of this scenario we clearly distinguish three phases of evolution: external helium-rich detonation, core compressive heating, and a final core carbon burn.
Though final disruption of the whole system is achieved at all resolutions, only models with minimum resolutions of 4~km and better exhibit all three phases. Particularly, core compression detonation is only observed for higher resolutions, producing qualitatively different nucleosynthetic outcomes. We identify the effect of finer spatial resolution on the mixing of hot silicon at the interface between the detonating helium layer and the underlying C/O WD as a primary driver of these dynamic differences. 
\end{abstract}

\section{Introduction}\label{sec:intro}
\replaced{}{\explain{Thorough revision of introduction. Changes to aim and focus of the text including removal of table 1 (cf. first submission), addition of specific examples, and a more straightforward discussion of numerical frameworks.}}
\edit1{Type Ia supernovae (SNIa) are among the most highly luminous of stellar transients, occurring in both old and young stellar populations \citep{2014ARA&A..52..107M}. Their number, distribution, and a tight relation between their peak luminosity and brightness decay have made them useful for cosmological scale distance measurements (\cite{1999ApJ...517..565P}, \cite{1998AJ....116.1009R}, \cite{2006A&A...447...31A}). They also play an important part in the chemical enrichment of galaxies (\cite{2007MNRAS.382.1050T}, \cite{2010MNRAS.401.1670F}, \cite{2020ApJ...895..138K}). Given their importance to these and other open questions, it is frustrating that our understanding of the details of the explosion mechanism is still incomplete. The production of large amounts of \ce{^{56}Ni} has been known to power their lightcurves for half a century at this point \citep{1962PhDT........25P, 1969ApJ...157..623C}, but details of the progenitor systems and the particulars of the nuclear burning that produce the radioactive nickel are a matter of active inquiry today.}

\edit1{In addition to the aforementioned production of \ce{^{56}Ni}, Type Ia supernovae are distinguished by the lack of H and He and the presence of distinct Si, Ca, and Fe features in their spectra. These spectroscopic features and the timescales over which they evolve have led to a strong consensus that their origin is the thermonuclear explosion of a C-O white dwarf \citep{2014ARA&A..52..107M}. Several scenarios have been invoked to explain the details of the explosion mechanism, including models with Chandrasekhar-mass white dwarfs (M$_{Ch}$) accreting material from both degenerate and non-degenerate companions and involving nuclear burning occurring in both subsonic (deflagrations) and supersonic (detonation) modes. However, in recent years, considerable attention has been paid to models of sub-Chandrasekhar-mass (sub-M$_{Ch}$) systems as potential progenitors that can produce results consistent with observations.}

\edit1{Here we focus on one specific model of this type: the double-detonation scenario. This scenario \citep{1982ApJ...253..798N, 1982ApJ...257..780N, 1994ApJ...423..371W}, involves accretion of helium from a rarefied donor onto the surface of a sub-M$_{Ch}$ white dwarf. A detonation in the accreted shell leads to the disruption of the white dwarf either through an {\it edge-lit} ignition of the underlying C-O core \citep{1982ApJ...253..798N, 1982ApJ...257..780N} or via the inward propagation of a shock into the core that eventually compresses a region of the core sufficiently to start a secondary {\it core detonation} \citep{1990ApJ...354L..53L, 1994ApJ...423..371W, 2014ApJ...785...61S}.
The large amount (roughly 0.1 \msun) of helium that was originally believed to be necessary for this model to work \citep[see, e.g.][]{1990ApJ...354L..53L, 1994ApJ...423..371W, 1996ApJ...457..500H} typically resulted in models with a significant overproduction of iron-group elements \citep{1996ApJ...457..500H, Nugent_1997}.
This led to studies of the viability of less massive shells \citep{Bildsten_2007, 2014ApJ...797...46S} and \citet{2010A&A...514A..53F} did realize core detonations in two-dimensional models with shells with masses less than 0.1 \msun.~\citet{2013ApJ...774..137M} also produced explosions in three-dimensional models using different ignition starting points and sizes, and helium shells enriched with some heavier elements.
These models produce reasonable candidates for the lower-luminosity edge of SNIa \citep{2014ApJ...797...46S}.}
\edit1{Some recent work \citep{2018ApJ...854...52S} has been centered on binary white dwarfs as progenitor systems for the double-detonation scenario. This scenario (the so-called 'D6' scenario) has been shown to be robust for a range of configurations while producing observational signatures that are roughly consistent with nucleosynthetic yields and ejecta velocities.}\par

\edit1{More recent studies have included the use of Lagrangian codes: \citet{2018ApJ...868...90T, 2019ApJ...885..103T} studied close double white dwarf binaries and the effects of the detonation on the donor star. \citet{2020A&A...635A.169G} investigated similar models to \citet{2010A&A...514A..53F}, but in three dimensions using a Lagrangian grid, a relaxation phase which mixes some of the shell and envelope, and somewhat coarser effective resolution. \citet{2020A&A...635A.169G} also implemented a nuclear network to model the energy release and isotopic composition of the burned material.  Core detonation is realized for all nine of their considered model systems, which include varying the ignition spot, using different masses, and varying the amount of carbon mixing into the shell. \citet{2019ApJ...878L..38T} exhibited similar results using an Eulerian code. \citet{2019ApJ...873...84P} performed a parameter study at around the similar resolutions, but in one spatial dimension. Importantly, for some combinations of shell and core mass and initial detonation position, \citet{2019ApJ...873...84P} note that edge-lit detonations result. They found that varying the ignition point position could instead produce a core detonation for the same shell and core masses and spatial resolutions.}

\edit1{This variety of edge-lit and core detonation outcomes and the seeming sensitivity to initial conditions motivates our work here. The differences between these two double-detonation sub-types could alter both the final composition of the remnant and the spatial distribution of that composition \citep{2019ApJ...877..136F}.  A source of uncertainty that has not been investigated systematically through parameter studies is the differing evolution of the detonation front at varying spatial resolution.
To this end, we investigate the hydrodynamic evolution of a double-detonation through over an order of magnitude in spatial resolution. We find that spatial resolution, in the context of an Eulerian, adaptive-mesh-refinement code without additional methods to control physical and numerical diffusion, does qualitatively impact the particulars of double-detonation explosion models, including  the  dynamics and other primary outcomes.}\par
This paper is structured as follows: Our methods and setup are described in Section \ref{methodology}. Section \ref{discussion} presents results of the simulations and a description of the observed phases of the explosion, whose duration and importance are functions of the grid resolution. In  Section \ref{conclusions} we discuss the implications of our results to future simulations of this scenario, including those in three spatial dimensions.

\section{Methodology}\label{methodology}
All of the simulations we describe here were carried out using the multiphysics code Flash-X \citep{DUBEY2009512, Harris2021}, an evolution of the widely-used FLASH code \citep{2000ApJS..131..273F} for exascale computer systems.
This modular code solves the compressible Euler's equations on a  Eulerian grid via a finite volume method. 
We also employ a small nuclear network (approx13, \citet{2000ApJS..129..377T}) to compute the energy release and change in composition of the stellar material as the detonation propagates .
The system of equations we must solve is completed using the Helmholtz tabulated equation of state \citep{2000ApJS..126..501T}.
Finally, the effect of gravity is computed via a multipole solver using 8 moments \citep{2013ApJ...778..181C}.

In addition to these particular choices of included physical modules, Flash-X uses adaptive grid refinement (AMR) to produce higher resolutions in key areas while maintaining a lower computational load elsewhere. In the simulations described here, refinement is triggered by tracking gradients in density, pressure, and molar fractions of the fluid. The resolution of density gradients are used because they mark the important features of the initial configuration, including the edge of the underlying white dwarf. Refining on pressure gradients allows the model to accurately track the position of the detonation shock and the compression wave that is observed in the stellar core. Finally, gradients in the mass fractions pinpoint the details of initial setup in a manner that guarantees a higher resolution around the shell-core interface from the start of the simulation. Additionally, 2 more refinement levels are allowed where energy release ($\epsilon_{nuc}$) is sufficiently large ($|\epsilon_{nuc}|>10^{16} \si{\erg\per\second}$), yielding the maximum refinement level only where burning is vigorously occurring (meaning its rate is enough to alter the dynamics of the simulation).\added{Burning within cells that fall within the numerically thickened shock is not allowed, as per \citet{1989nuas.conf..100F} and \citet{2014ApJ...782...12P} to prevent numerical diffusion from affecting the propagation of the reaction front. \citet{Townsley_2016} have shown that this practice reproduces the steady-state detonation speed and does not affect the extended reaction structure beyond the initial carbon burning stage in the FLASH code.}

We note that burning is moderated via the Flash-X runtime parameter \texttt{enucDtfactor} ($\varepsilon$) which restricts the time step to be a constant fraction  of the \replaced{nuclear energy release timescale, calculated as the relative change of internal energy over the hydrodynamical timestep $\left(\varepsilon\frac{\delta E_B}{\delta t}\right)$.}{nuclear energy release timescale, calculated as the ratio of the internal energy ($E_B$) and the nuclear energy from burning per unit time ($\epsilon_{nuc}$) over the hydrodynamical timestep $\left(\varepsilon\frac{E_B}{\epsilon_{nuc}}\right)$.} \replaced{Recently, new methods of restricting the energy release from nuclear burning have been suggested, namely \citet{2021ApJ...919..126B} implementing the limiter outlined in \citet{2013ApJ...778L..37K}.
This technique evolves the simulation at the CFL-limited time step, relaxing the restriction imposed from fully resolving the nuclear burning timescale. This prevents sudden changes in the nuclear energy release at the onset of a detonation in the context of an operator-split scheme from sharply reducing the overall time step by throttling the nuclear energy release and concomitant composition changes. These types of limiters, though keeping energy deposited in the simulation in check, may directly affect the dynamics of the detonation and therefore the final yields.}
{A method of restricting the energy release from nuclear burning has been suggested by \citet{2013ApJ...778L..37K}, and this scheme has been used in several studies with physical parameters like the ones described here \citep[e.g.,][]{2021ApJ...919..126B}. 
This technique evolves the simulation at the CFL-limited time step, relaxing the restriction imposed from fully resolving the nuclear burning timescale by renormalizing all of the burning rates within a cell if the energy delivered in a time step exceeds some fraction of the cell's internal energy. This prevents sudden changes in the nuclear energy release at the onset of a detonation in the context of an operator-split scheme from sharply reducing the overall time step. \citet{2019ApJ...874..169K} caution that the use of such a limiter may lead to unexpected suppression of burning in regions where ignition should occur.  However, \citet{Kushnir_2019} argue that this finding and many of the details in \citet{2019ApJ...874..169K} are not generic to the use of the limiter, but, rather, depend on specifics of the simulations carried out therein. Because we do not require the use of the limiter here and our desire is to restrict our study only to the particular effects of spatial resolution in our Eulerian code, the limiter of \citet{2013ApJ...778L..37K} is not implemented in our simulations.\explain{expanded burning limiter discussion}}

\begin{figure*} 
\begin{center} 
\includegraphics[width=\textwidth]{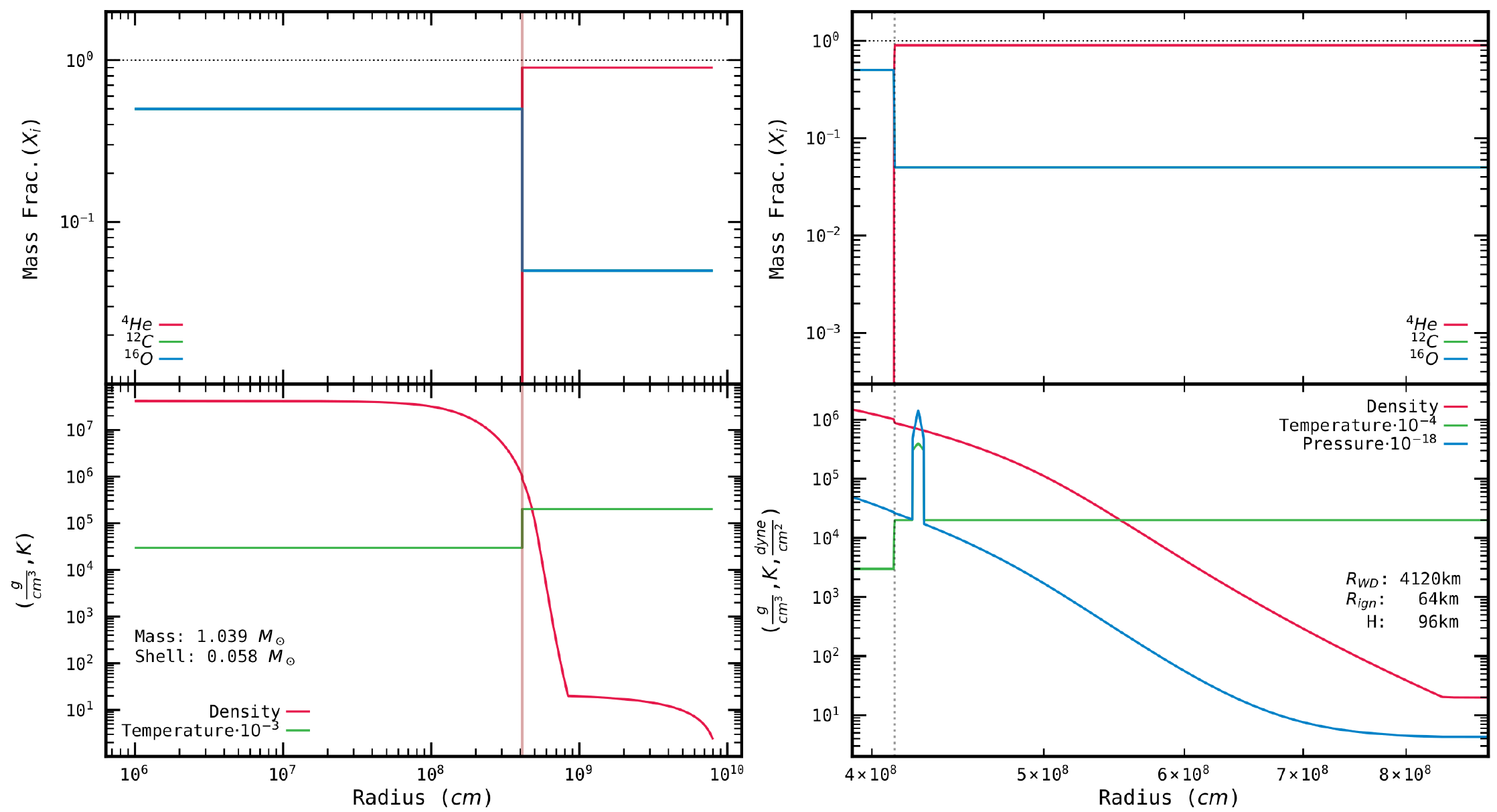}
\end{center}
\caption{Left: Spherically symmetric, hydrostatic analytical profile made with initial conditions for a central density and composition and closed with the Helmholtz equation of state. Right: Magnified line-out of the initial profile above the WD surface in Flash-X with the heated bubble added to ignite the detonation.}
\label{fig:figure1}
\end{figure*}

\subsection{Numerical Setup}\label{sec:numerical_setup}
The domain uses cylindrical symmetry and is sized to encompass both the white dwarf itself and its immediate surroundings, reaching an extent of $\{r, z\}=\{\SI{6.5536e4}{\kilo\m},\SI{13.1072e4}{\kilo\m}\}$. This domain size is required (indeed, it is likely the minimum useful extent) because the envelope is expelled considerably outwards before the onset of any core detonation. At the same time, upper limits for the domain size must be chosen so as to avoid the accretion disk and inclusion of the companion, given our ignorance of how the accretion process actually impacts the dynamics. 
The burning module uses Flash-X's approx13 13-species network with \ce{^{4}He}, \ce{^{12}C}, \ce{^{16}O}, \ce{^{20}Ne}, \ce{^{24}Mg}, \ce{^{28}Si}, \ce{^{32}S}, \ce{^{36}Ar}, \ce{^{40}Ca}, \ce{^{44}Ti}, \ce{^{48}V}, \ce{^{52}Cr} and \ce{^{56}Ni}. This network uses effective rates tuned to more accurately capture nuclear energy generation and captures the bulk behaviors of isotopic flow and hydrodynamic feedback for the thermodynamic conditions and timescales encountered in the simulations we considered.

The initial 1D profile is based on Model 3 from \citet{2010A&A...514A..53F}. We construct the model by directly querying Flash-X's equation of state with an initial central density of \SI{4.15e7}{\g\per\cubic\centi\m} and a composition for the white dwarf of $X_{\ce{^{12}C}}=X_{\ce{^{16}O}}=0.5$.
This is integrated outward subject to hydrostatic equilibrium and assuming spherical symmetry to a density of $10^6$ \si{\g\per\cubic\centi\m}, \replaced{where the composition is changed to the aforementioned mixed value for the shell. The shell continues to densities of roughly \SI{20}{\g\per\cubic\centi\m}.}{where the composition is changed to a mixed value of $X_{\ce{^{4}He}}=0.90, X_{\ce{^{12}C}}=X_{\ce{^{16}O}}=0.05$ to start the shell. Outward integration continues until the shell reaches densities of roughly \SI{20}{\g\per\cubic\centi\m}.\explain{not "aforementioned", only shown in figure 1.}}
Finally, for numerical stability purposes, a hot, exponential density scale height tail is added above the shell extending to the domain limits. 
This tail is read into the initial conditions in Flash-X as an isothermal fluff.

\replaced{A total of four simulations were performed at maximum resolutions of \SI{16}{\km}, \SI{8}{\km}, \SI{2}{\km}, and \SI{0.5}{\km}.}
{A total of five simulations were performed at maximum resolutions of \SI{16}{\km}, \SI{8}{\km}, \SI{4}{\km}, \SI{2}{\km}, and \SI{0.5}{\km}.\explain{erratum. There are five simulations total}} These resolutions represent the highest levels of grid refinement reached by the AMR and cover salient regions of burning in the domain.
\replaced{All 4 simulations}{All five simulations\explain{see above}} are initialized with the analytical profile shown in Figure \ref{fig:figure1}, which amounts to a C-O core mass of \SI{0.981}{M_\odot} 
\replaced{with an overlying \ce{^{4}He}-shell mixed with 5\% \ce{^{12}C} and 5\% \ce{^{16}O} by mass fraction. There is an additional \SI{0.058}{M_\odot} of \ce{^{4}He} in the low-density ``fluff'' filling the rest of the domain.}
{with an overlying \ce{^{4}He}-shell mixed with 5\% \ce{^{12}C} and 5\% \ce{^{16}O} by mass fraction adding a total additional \SI{0.058}{M_\odot} to the domain. This envelope split into \SI{0.050}{M_\odot} forming the shell proper out to \SI{8410}{\kilo\m}, and \SI{0.008}{M_\odot} in the low-density ``fluff'' filling the rest of the domain.\explain{clarification. pollutant is mixed in the whole envelope. Separation of shell mass and fluff edge mass.}} The assumed carbon-oxygen pollution in the envelope stems from multiple mechanisms before ignition: the companion's accretion stream composition, burning of some helium in the shell, and dredge-up from the core into the envelope \citep{2013ApJ...776...97M, 2014ApJ...797...46S}. \added{Our choice for initial enrichment of the fuel(5\% \ce{^{12}C} and 5\% \ce{^{16}O}) is motivated primarily from the results of \citet{2013ApJ...776...97M}, where this choice produced detonation velocities and burning length scales that were roughly the median of the values considered therein.\explain{pollutant range.}}

To achieve ignition in the accreted layer, sufficient energy must be deposited in a small enough area to produce a self-sustaining detonation in the helium-rich shell.
This is done by directly altering the thermodynamic state of a selected set of zones in the shell.
An increased temperature at a given density is typically used to start detonations of this type, but this does not guarantee the detonation will be self-sustaining.
Therefore, the spatial extent of the heated region and the gradients in various values must also be chosen carefully.
\replaced{\citet{2013ApJ...774..137M} achieve self-sustaining detonations in both 2D axisymmetric models with maximal resolution of \SI{6}{\kilo\m} by increasing the temperature in zones within a radius of \SI{25}{\kilo\m} from the ignition point using a peak ignition temperature of \SI{2}{\giga\K} that has a slight linear decline towards the edge of the region. This prescription is used in the least massive of their considered shells (0.072 $M_\odot$) that are composed of pure helium.}{\citet{2013ApJ...774..137M} achieve self-sustaining detonations in their least massive, pure-helium shell (0.045 $M_\odot$), 2D axisymmetric model by increasing the temperature in zones within a radius of \SI{25}{\kilo\m} from the ignition point using a peak ignition temperature of \SI{2}{\giga\K} that has a slight linear decline towards the edge of the region.\explain{erratum: most comparable simulation has a smaller shell than previously noted.}}\added{Notably, \citet{2013ApJ...774..137M} achieve detonation propagation  only when the resolution reaches \SI{2}{\kilo\m}, with coarser resolutions not able to sustain the detonation.}
\replaced{\citet{2018MNRAS.476.2238G} include mixing of carbon and oxygen into the shell and posit a convective timescale that can easily reach the starting temperatures for detonation. This happens a few hundred kilometer above the core-shell interface.}{\citet{2018MNRAS.476.2238G} model convection of accreting helium onto the core with a 2D wedge centered around the core-shell interface. While convection mixes the helium with the carbon-oxygen core blurring the interface, detonation conditions (burning timescales faster than both hydrodynamical and convective timescales) are achieved at around \SI{100}{\kilo\meter} above the closest pure carbon-oxygen composition boundary.\explain{rewritten sentence to make clearer the impact of Glasner et al. (2018) on our initial detonation position}}\added{Finally, \citet{2010A&A...514A..53F}, do not achieve a detonation for a small (of order a few \SI{8}{\kilo\m} cells) \SI{3}{\giga\K} ignition zone.} 
Considering these previous studies, we set the initial ignition condition as a heated bubble at the pole, centered at \SI{100}{\kilo\m} above the surface of the C-O core. This region has a radius of \SI{32}{\kilo\m}, regardless of resolution, to achieve a relevant size in the \SI{16}{\kilo\m} case. Within this region, temperature is set to \SI{3}{\giga\K} and linearly increases towards the center of the bubble to \SI{4}{\giga\K} \replaced{to begin a detonation in the shell}{to ensure a detonation in the shell at all tried resolutions\explain{clarity on picked temperature.}}.
A line-out through the central ignition point of this 2D initial configuration is shown in the right panel of Figure \ref{fig:figure1}.

\section{Results}\label{discussion}
\subsection{Dynamics}
\begin{figure*}  
\begin{interactive}{animation}{figure2.mp4}
\includegraphics[width=\textwidth]{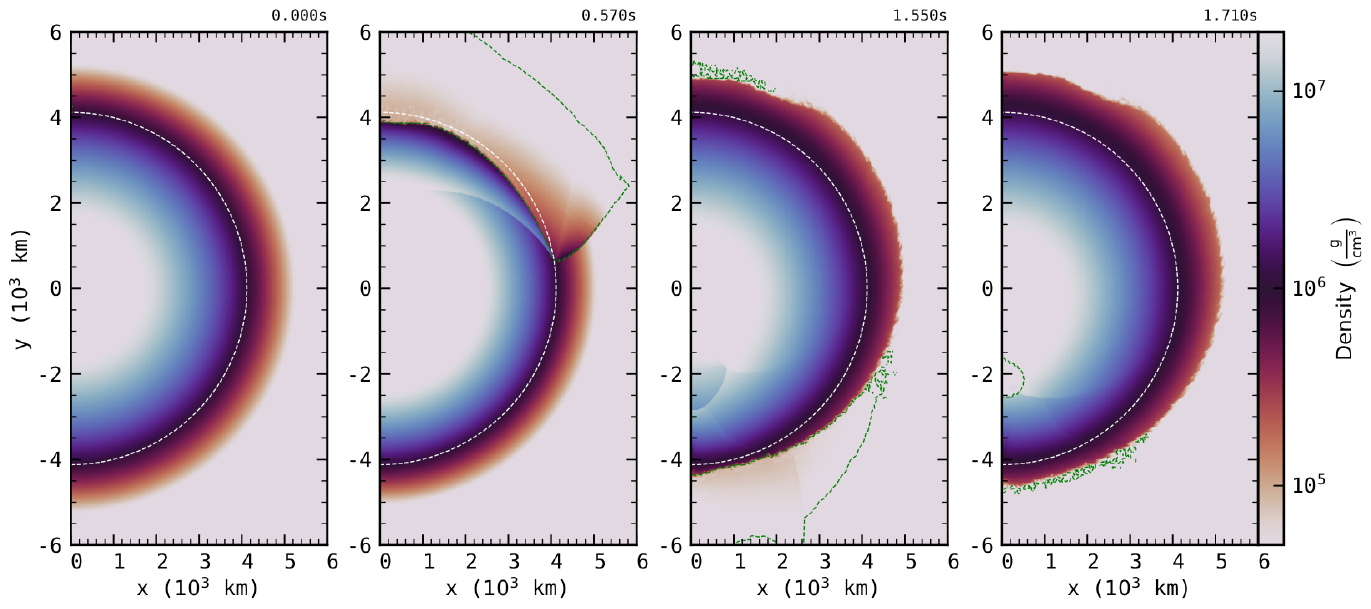}
\end{interactive}
\caption{Key features in the evolution in density for the maximum resolution case (\SI{0.5}{\kilo\m}). The white contour marks the initial edge of the WD (\SI{4120}{\kilo\m}), while the green contour traces a temperature of \SI{1}{\giga\K}. To start the detonation, a small region is artificially heated above the WD along the axis of symmetry as described in Section~\ref{sec:numerical_setup}. Initially, a supersonic front propagates around the edge of the core, burning mostly shell material with some incursion into the core via mixing. After reaching the antipodal point at 1.2~s, shell burning subsides. Throughout this stage, the core is compressed due to the initial shock of the ignition aided by the surrounding burning front. These compression waves converge at an off-center point inside the core at around 1.6~s, causing a secondary carbon detonation that unbinds the system. A single panel animation of the time evolution is available. The animation shows the density, covers the same spatial extent and runs from 0 to 2.6 s highlighting the same contours as shown in the figure.}
\label{fig:figure2}
\end{figure*}

For all resolutions tested here, the entire star is unbound within three seconds of the initial helium-shell ignition, however the process of the explosion depends on resolution. The principal features of the highest resolution scenario are shown in figure \ref{fig:figure2}. The initial edge-lit detonation is not strong enough to ignite the core outright, but rather the detonation front makes its way around the shell, quickly enveloping the C-O core and generating mostly intermediate mass elements as products of explosive helium burning.
The detonation propagates along the limb of the core (shown as the green contour in Figure \ref{fig:figure2}), increasing the lateral pressure and creating a pressure wave directed inward, but not strictly radially. After $\approx$1.5s, the shell, the initial innermost extent of which is marked by the white contour in Figure \ref{fig:figure2}, has expanded sufficiently to quench any relevant composition changes and the shell detonation subsides.  Shortly after this time, the initial inward pressure wave converges at an off-center point in the core, igniting a second detonation via compressional heating. This detonation is strong enough to unbind the white dwarf, as shown in the last panel of figure \ref{fig:figure2}. We refer to this outcome as Case 1 and note that it only obtains in simulations with minimum cell widths of less than or equal to \SI{2}{\kilo\m}. \par

\begin{figure*}  
\begin{center}
\includegraphics[width=0.5\textwidth]{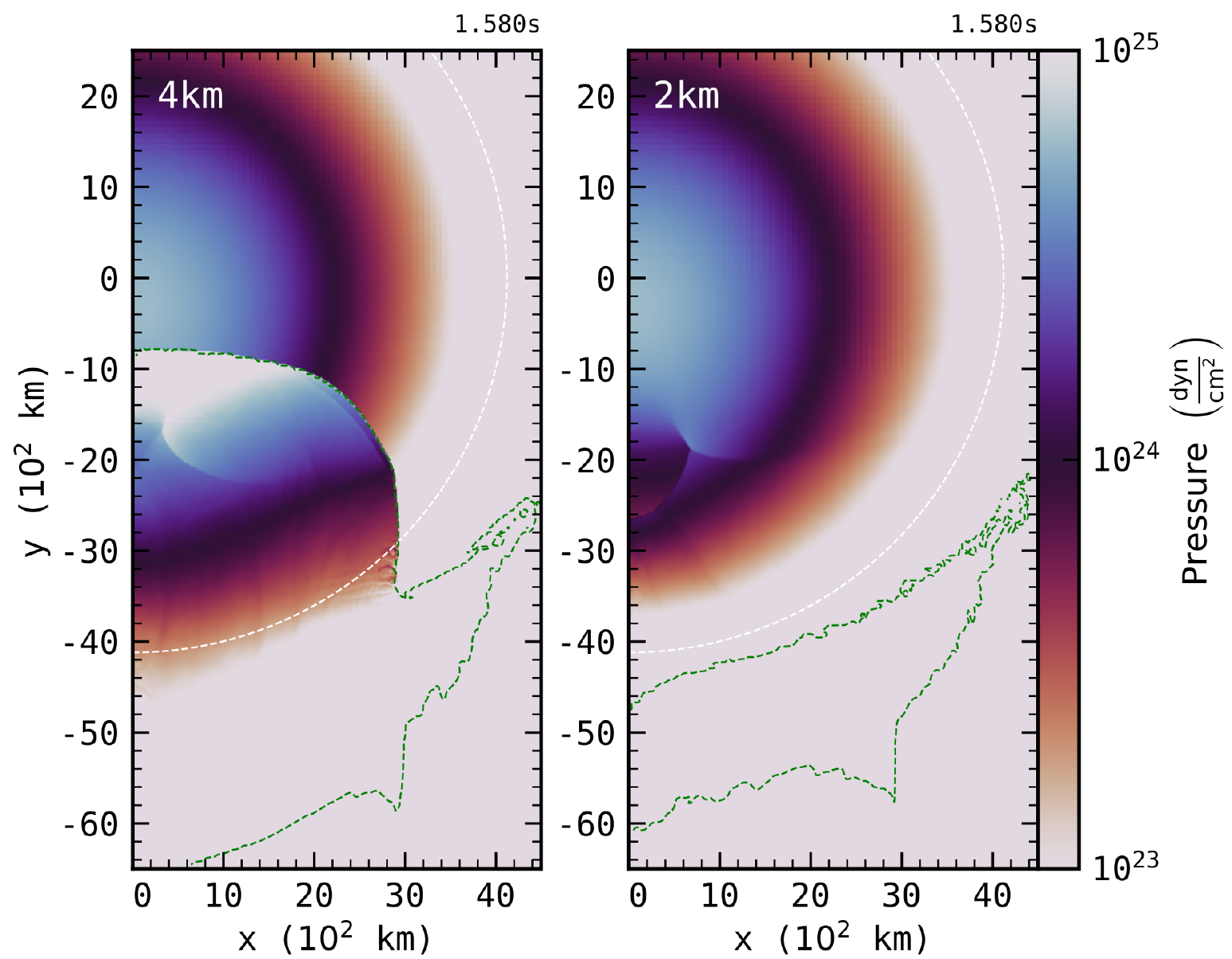}
\end{center}
\caption{A magnified view of the antipodal point for two different maximum resolutions. As in the previous figure, the white contour marks the WD extent while the green contour demarcates a temperature of \SI{1}{\giga\K}. Left: The four-kilometer resolution case is driven not by a core ignition but rather an antipodal detonation. In this case, shell burning continuously mixes and progressively burns some mixed core material. Right: The two-kilometer case driven by off-center compression. This case burns the core after a distinct pause in shell burning, and separates the two detonations spatially and therefore also in terms of density.}
\label{fig:figure3}
\end{figure*}

\begin{figure*}  
\begin{center}
\includegraphics[width=0.5\textwidth]{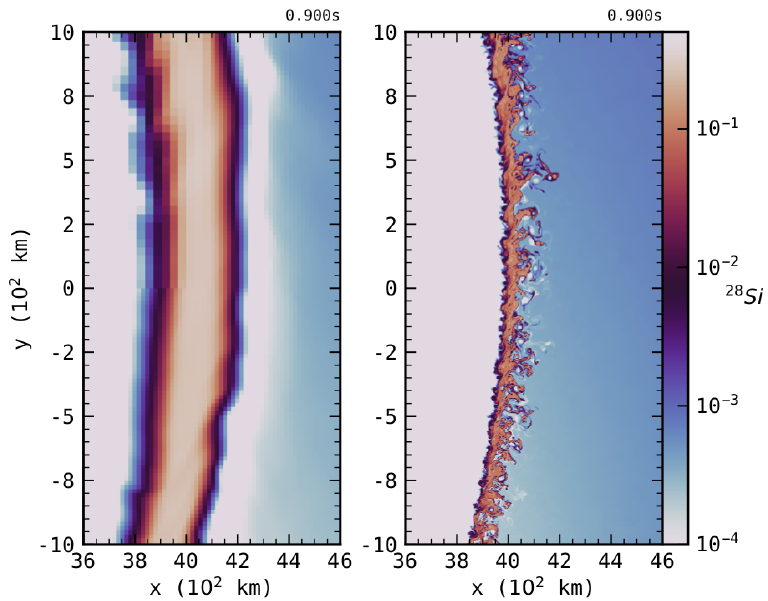}
\end{center}
\caption{Outer shell burning remnants: equatorial slice of mass fraction of silicon-28 at the WD limb after shell burning traverses the whole cutout. At 16~km resolution (left), the thickness of the expanding band is 4 times larger than in the highest resolution case (0.5~km, right). Additionally, mixing is far more complex and evolved even at scales which do not represent the actual length scales of burning.}
\label{fig:figure4}
\end{figure*}

\begin{figure*}  
\begin{center}
\includegraphics[width=\textwidth]{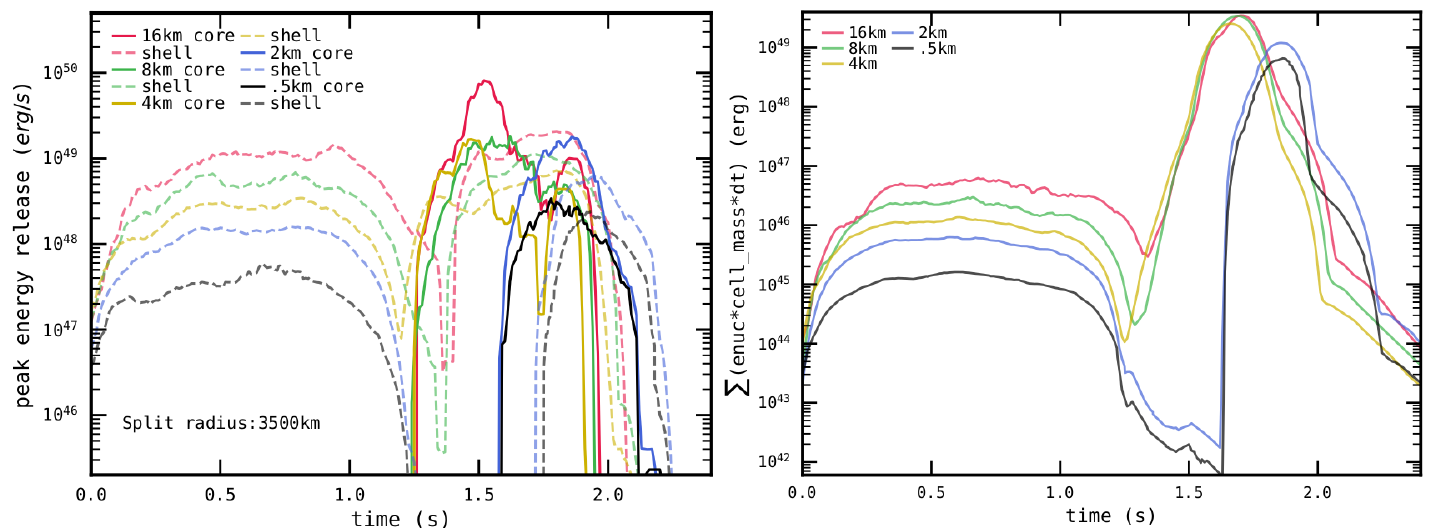}
\end{center}
\caption{Left: Peak positive energy release within the domain, split into values for the 'core' and 'shell' based on radial distance from the domain center (split radius). Right: Total energy added to the domain via burning. Two families of solutions emerge based on the dynamics of the event: a direct shell to core burning at resolutions over 4~km, and a case with clearly separated shell and core burning stages at finer maximum resolutions.}
\label{fig:energies}
\end{figure*}

A pivotal difference appears at resolutions of \SI{4}{\kilo\m} and above, as shown in figure \ref{fig:figure3}. The wake of burned material behind the detonation front deposits a threshold amount of energy at a point on the limb of the WD when the front reaches the symmetry axis, causing a secondary antipodal (with respect to the original ignition point) detonation, which ignites the core at its edge. This detonation immediately propagates into the core. It obviates the effect of the compressional heating from the earlier stages of burning by burning through the convergence point before the pressure wave reaches that locus (Case 2). This is in contrast to higher-resolution runs, where the shell burning is quenched by the expansion of the envelope and does not produce a secondary shell detonation. In these higher-resolution simulations, the only process producing a core detonation is a separate ignition through compression (Case 1, as described above). We once again note that due to the asymmetric ignition condition, the convergence point is far from the geometric center of the WD; namely, over \SI{2000}{\kilo\m} from the center for all examined resolutions.\par

\begin{figure} 
\begin{center}
\includegraphics[width=0.4\textwidth]{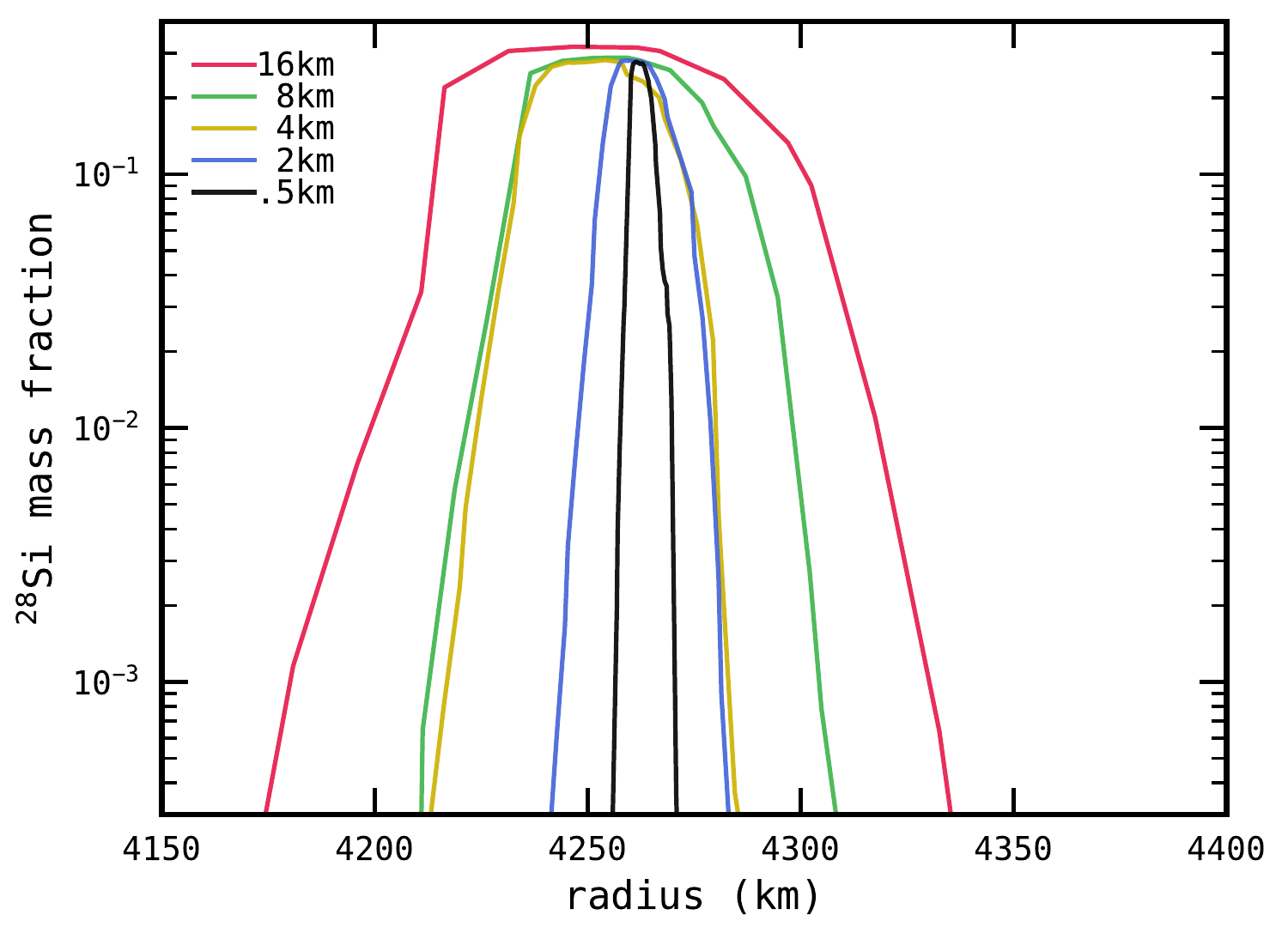}
\caption{Silicon burning front thickness for all cases, measured at 20 degrees from the antipode.}\label{fig:figure6}
\end{center}
\end{figure}

The core-shell-interface detonation observed in the lower-resolution cases is driven by the compression of hot, partially burned and significantly mixed material as the initial shell detonation front collides with itself at the reflecting boundary at the second pole of the configuration. This material has been swept up during the earlier progress of the detonation front around the edge of the WD. The coarse resolution of the less-resolved cases produce a mass of compressed material at this focus point that is considerably large than in the higher-resolution cases because of the relative lack of spatial resolution (see figures \ref{fig:figure4} and \ref{fig:figure6}). In addition, this material is preconditioned for explosion via the artificial mixing of hot, partially burned shell material and underlying core material. This numerical effect is apparent in Figure \ref{fig:figure4}. The coarser resolution spuriously smooths sharp gradients in composition at the shell-core interface, causing the suppression of fluid instabilities that are clearly visible in the higher-resolution cases. Without this additional numerical mixing, the antipodal material is insufficiently compressed and heated by encountering the reflecting boundary at the second pole, allowing time for the off-center core detonation to be triggered deeper in the core of the WD. 

More quantitatively, we calculate the internal energy of the matter for the domain within \SI{4e8}{\centi\m} of the center of the system (the 'core') and outside of said radius (the 'shell'). Both the peak energy release as a function of position and the total overall energy being deposited in the matter are shown in figure \ref{fig:energies}. The left figure shows the explicit separation of  shell and core burning and the transition in the nature of this distinction as resolution is increased. Evident is a distinct shift of approximately 0.5 s in core ignition for case 1. The right figure further  reinforces this separation by showing that at resolutions finer than \SI{4}{\kilo\m}, burning subsides in the shell before the core can be ignited at the limb by the original detonation wave.
\added{In all of the simulations we find that the detonation speeds during the burning of the envelope agrees well with other simulations at around \SI{10500}{\kilo\m\per\s}\citep{2010A&A...514A..53F, 2020A&A...635A.169G}. We also find consistent detonation speeds in simpler configurations (i.e. two-dimensional burning tubes) using the same reactive flow code, also agreeing with earlier work \citep{2012ApJ...755....4T}. The curvature of the propagating front in all simulations described here tends to \SI{0.05E-7}{\per\cm}, which according to the analytical solutions given by \citet{2013MNRAS.431.3429D} for similar densities and a pure helium fuel correspond to speeds in a range of \SIrange{10000}{12000}{\kilo\m\per\s}.}\par

\added{We note that previous investigators have also obtained a variety of outcomes that are qualitatively similar to our Cases 1 and 2. \citet{2010A&A...514A..53F} used a level set with two advected scalars to model burning in their simulations, effectively preventing core burning until the core reaches a certain temperature. Their results are very much like our Case 1:  off-center detonation due to converging waves from the initial shell detonation.  \citet{2013ApJ...774..137M} probed multiple initial detonation sites in both two and three dimensions, obtaining an off center hotspot in all simulations. More recently, \citet{2021ApJ...919..126B} running with a similar codebase in two dimensions obtained Case 1 scenarios for all but the heaviest shell they investigated ($0.1\msun$). That heaviest case is very similar in origin and evolution to our Case 2. \citet{2018ApJ...868...90T, 2019ApJ...885..103T}, modeling a large three dimensional domain that includes both progenitor and companion, also obtain outcomes very similar to our Case 1 and notably show a distinct thick silicon shell which maintains its shape well out into the domain.  
A notable counterpoint is found in the three-dimensional simulations of \citet{2020A&A...635A.169G}. In a setup similar to ours on a Lagrangian grid, they find a Case 2-like scenario with a distinct antipodal detonation (they dub this a ``scissor mechanism'') where the shell burning front crosses over itself at the antipode, pinching the partially mixed limb of the core enough to start a detonation. Following the pinch, this limb detonation reaches the expected core ignition point before a Case 1 scenario occurs. 
}

\subsection{Nucleosynthesis}
\begin{figure}  
\begin{center}
\includegraphics[width=0.4\textwidth]{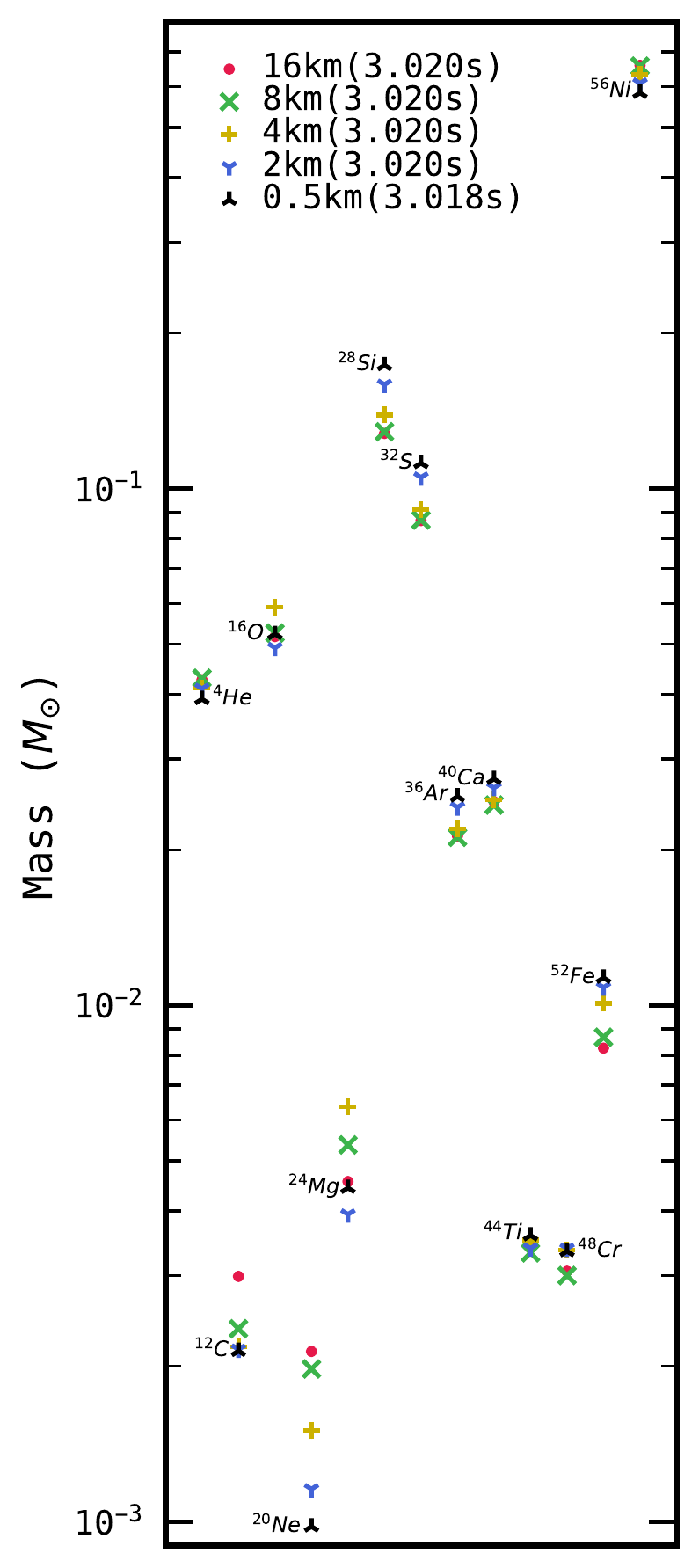}
\caption{Final yields from all simulations.}\label{fig:figure7}
\end{center}
\end{figure}

Final yields from the modeled explosions are shown in Figure \ref{fig:figure7}, taken at a time where the energy release falls by ten orders of magnitude from shell-burning levels.
Using the two higher-resolution cases as fiducial (2 and 0.5 \si{\kilo\m}), their relative differences in yields vary by less than 10\% except in 2 species: Mg(11\%) and Ne(18\%). 
With respect to differences that appear across both cases, larger and distinct differences are seen when comparing the \SI{4}{\kilo\m} vs the \SI{2}{\kilo\m} case. In this comparison,  the largest differences arise in magnesium(62\%), neon(30\%), silicon(13\%), and oxygen(20\%).

Shell burning is remarkably similar in all cases, regardless of resolution, particularly in terms of the velocity of the burning front 
(where the detonation velocity differs by no more than 2\% 
\replaced{over a mean of \SI{8244}{\km\per\s} among all the simulations}{)\explain{Deletion. Velocity deviation was measured from mean fluid velocity behind the shock, which is not representative of the detonation velocity but still relevant in terms of deviation among resolutions.}}. However, as previously discussed, there is significant variation in the thickness of the mixing layer between the burning shell and the underlying core, with coarser resolutions mixing the edge of the core as the detonation goes around it. 
\replaced{These thicker fronts produce a higher mass of intermediate elements as opposed to higher resolution cases, which do not burn as much core material under shell burning conditions.
Since shell burning is the principal contributor of neon, the effect of the artificial thickening of the burning front due to resolution is best observed in this species (see figure \ref{fig:figure7}), thus thicker resolutions should leave a trace of this burning stage by yielding a relatively higher mass of intermediate elements.}{These thicker fronts produce a higher mass of intermediate elements as opposed to higher resolution cases, which do not mix and burn as much core material under shell burning conditions (Figure 7). Since shell burning is the principal contributor of neon, the effect of the artificial thickening of the burning front due to resolution is best observed in this species, where the difference in yield from our coarsest to finest resolution cases is roughly a factor of two.}

Other intermediate species (Mg, Si, S) show a large variation with resolution, not only in their yields but also in their spatial extent (see figure \ref{fig:figure4}). This is a direct effect of improving resolution in a problem where the real burning front is never resolved, and results in a distinctly thinly layered composition as opposed to a thick, more thoroughly mixed composition.

\begin{figure}  
\begin{center}
\includegraphics[width=0.4\textwidth]{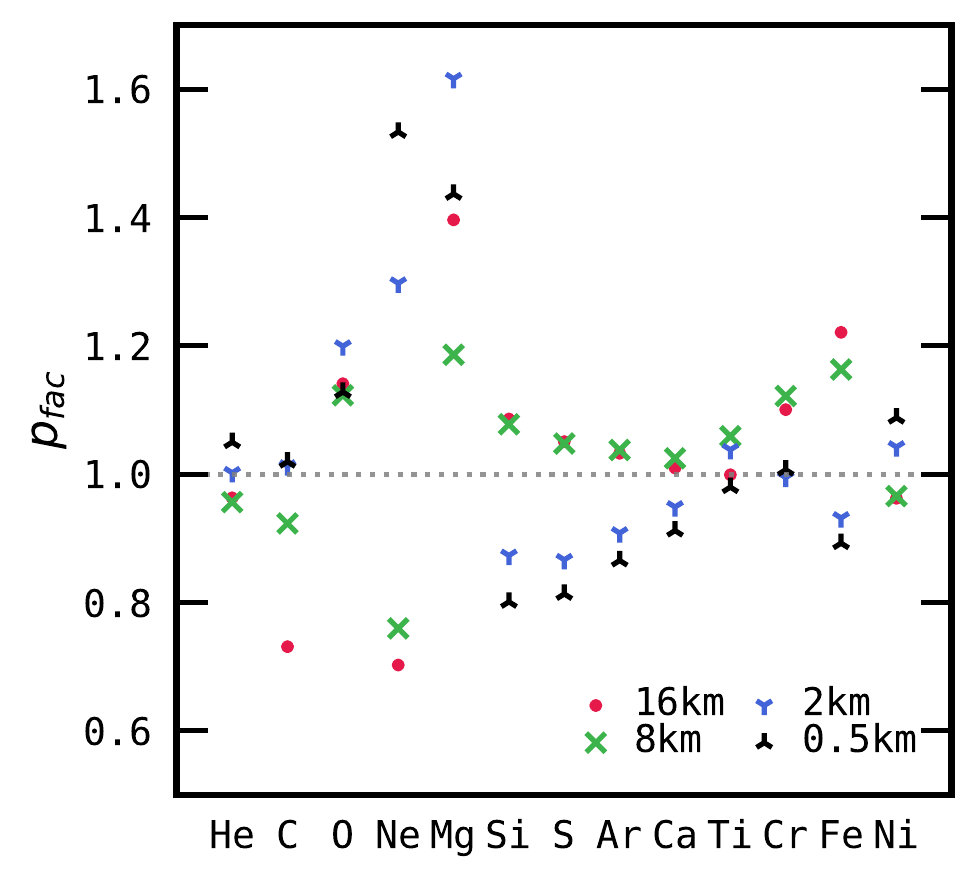}
\caption{Relative yields for simulations compared to the 4km case.}\label{fig:figure8}
\end{center}
\end{figure}

\begin{figure*}  
\begin{center}
\includegraphics[width=\textwidth]{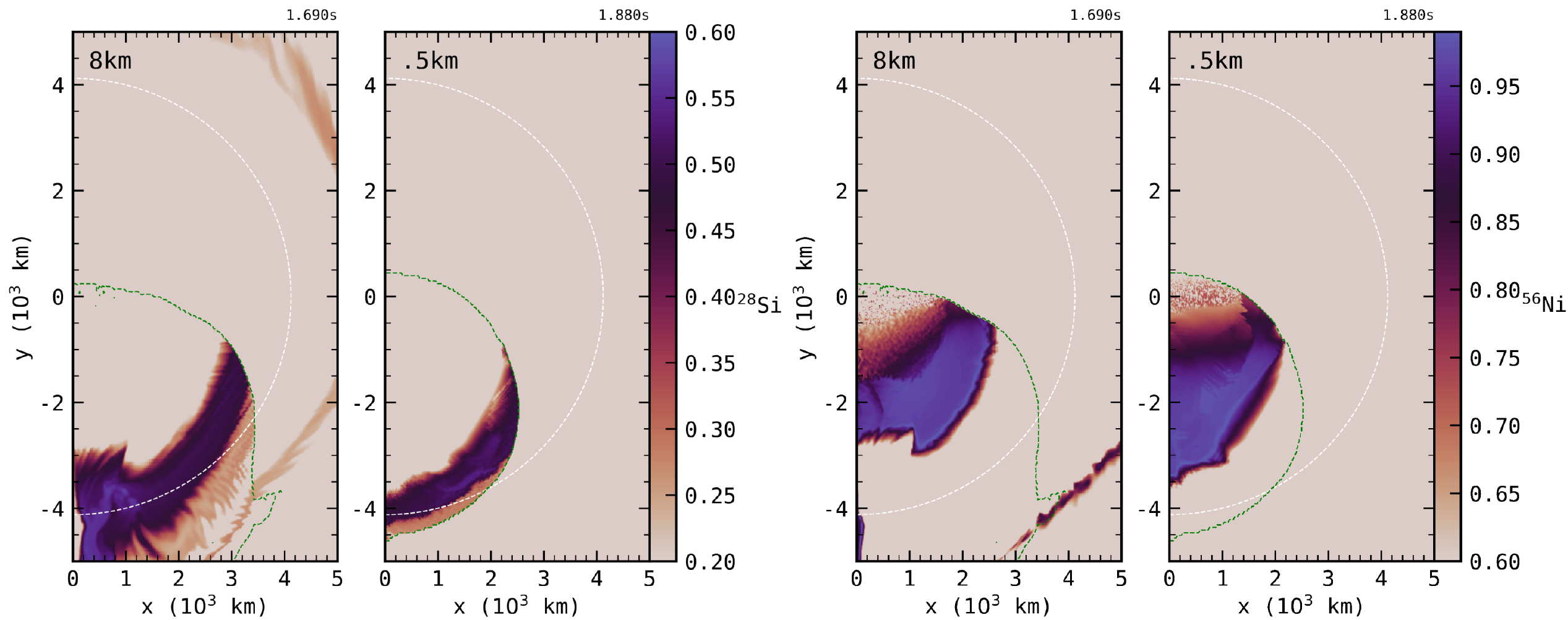}
\caption{Silicon and nickel compositions around the core at the time of peak energy release for both cases. Left: A large amount of low density material is burned to silicon at coarser resolutions due to the secondary detonation starting at the edge of the white dwarf. Right: Burning to nickel in coarser resolutions is localized to the front climbing the density gradient of the core, while for higher resolutions (case 1), the off-center detonation starts at a higher density resulting in more complete burning over the lower hemisphere of the core. Subsequently, both cases evolve similarly over the rest of the domain.}\label{fig:figure9}
\end{center}
\end{figure*}

Coarser resolutions burn from the limb of the WD, causing the core to burn at progressively higher densities towards the center and compressing the material closer to the center. On the other hand, finer resolutions manage to ignite the core from a near-central position, causing core burning to start at a higher density and progress towards the lower densities at the edge of the WD. This difference in the cases causes what is shown in figure \ref{fig:figure8}.
\replaced{In Case 1, the detonation starts at a lower density, and climbs the core density gradient, yielding comparatively less nickel. As for Case 2, the detonation starts at higher densities than the coarser resolution cases. The detonation propagates down a decreasing density gradient towards the limb of the WD. This raises the temperature of the front as it progresses and yields a higher total mass of nickel.}
{Further clarity can be seen in figure \ref{fig:figure9}. For Case 2, due to the secondary detonation starting at a lower density near the edge of the white dwarf, a substantial portion of the core is burned while the front is climbing the density gradient towards the center of the core. This material is burned at lower temperatures yielding more intermediate mass elements. This behavior is exhibited via the silicon mass fraction, shown in the left group of figure \ref{fig:figure9} for two representative resolutions. On the other hand, Case 1 begins its secondary detonation closer to the center of the core at a much higher density. This burning front starts at high temperatures and propagates outward, down the density gradient of the core. But, because the front starts inside the core, much more material (at higher density) is quickly burned to nickel. The front eventually reaches most of the material that Case 2 burned, but encounters this material at a higher temperature than was realized in Case 2, burning most of that material all the way to nickel (shown in the right panels of figure \ref{fig:figure9}). 
\explain{Typo: case numbers switched. Corrects confused explanation, spatial extent imprints this effect in a new figure (figure 9).}}

{In Case 2, the detonation starts at a lower density, and climbs the core density gradient, yielding comparatively less nickel. As for Case 1, the detonation starts at higher densities than the coarser resolution cases. 

\explain{Typo: case numbers switched.}}

\added{figure 9, spatial extent of IGE yields}

\subsection{Ejecta structure}\label{subsec:ejecta}
\begin{figure*}  
\begin{center}
\includegraphics[width=\textwidth]{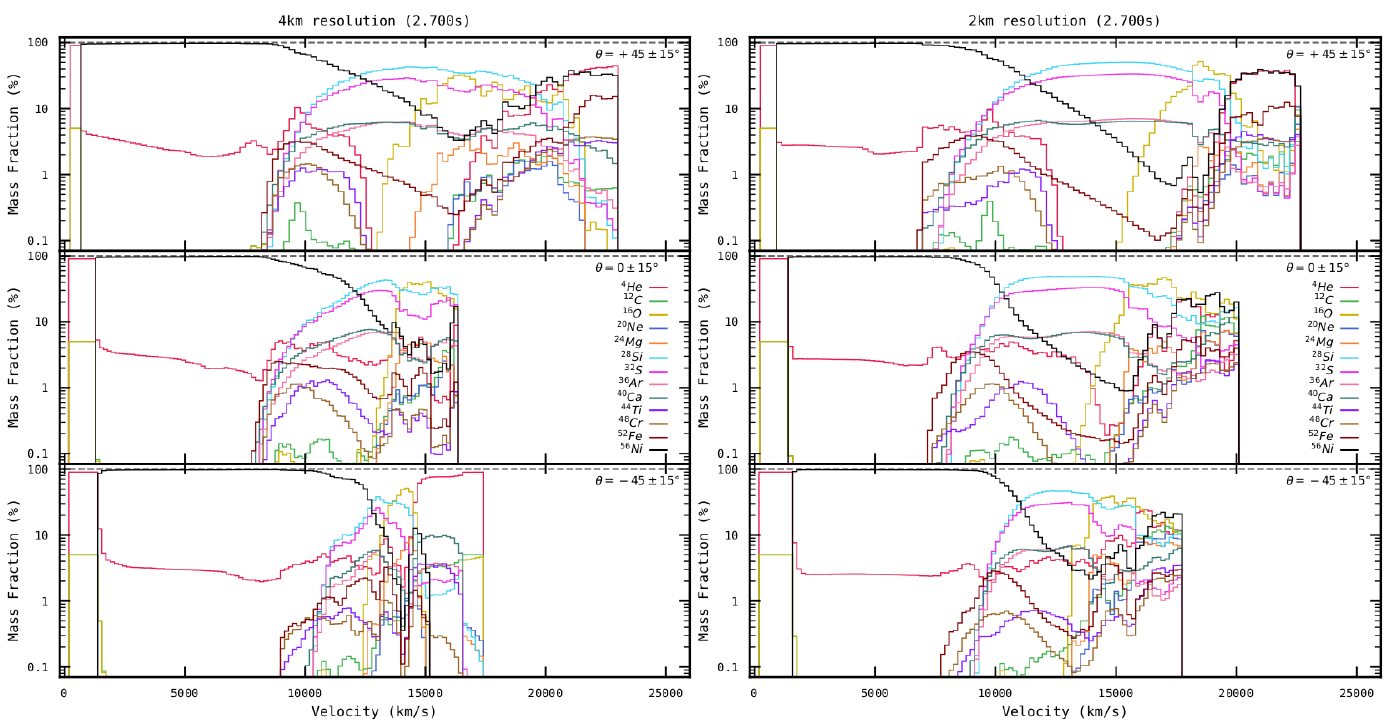}
\end{center}
\caption{Ejecta composition by mass fractions sliced by angle of view. Top, middle and bottom enclose 30 degree wide slices in the domain centered at noted angles ($\theta$\added{, further description in section \ref{subsec:ejecta}\explain{see above}}). Left: Described slices for the 4~km case. Right: Described slices for the 2~km case.}
\label{fig:figure10}
\end{figure*}

The final simulation ejecta structures are shown in Figure \ref{fig:figure10}.\added{Shown are velocity distributions of three 30 degree sized wedges fanning out into the domain from the geometrical center with their angular center aimed at 3 specific angles ($\theta$) each measured from the center of the core hemisphere ($\{x,y\}=\{0,0\}$, and defined such that positive angles point towards positive y): one pointing towards the equator ($\theta=0\degree$), one at 45 degrees ($\theta=45\degree$) north of the equator, and one at 45$\degree$ south of it.\explain{poor description in figure, expanded in text}} Though at the plotted times homology has not yet been reached, all burning has already subsided and most mixing has already taken place. Given this, some of the  principal characteristics and the overall structure of the ejecta can be seen at these early times and are not expected to change in character at later times. 

First, aside from the focusing and compressional effect of axisymmetry that was noted earlier, another consequence of the reflecting boundary along the polar axis can be discerned. \replaced{The reflecting boundary condition compresses material hitting it, leading to acceleration along the polar axis, causing a boost in velocities at the poles. Though this happens in both cases, the secondary surface detonation of Case 2 exacerbates this effect by having different compositions on each side of the detonation.
In Case 1, the off center detonation diminishes this effect by burning core material outward in all directions. There is also less material near the antipode in this case, bringing the antipodal velocities closer to the polar ones imparted by the initial shell detonation.}
{The reflecting boundary condition diverts material hitting it, pushing the material along the polar axis and causing a boost in velocities at the poles. Though this happens in both cases, the surface detonation of Case 2 imparts an additional acceleration to the diverted envelope material along the axis. This same sort of behavior is evident above and around the initial polar detonation. For Case 1, there's no additional energy release at the antipode, leading to lower polar velocities in this case. Additionally, the secondary detonation in Case 1 pushes against a significant overburden of core material between the ignition and the envelope, causing a comparably smaller boost at the antipode.\explain{clarified the effect of the BC's and the different cases}}
Second, though nuclear burning has subsided, there is still a significant amount of helium on the grid. Notably for Case 2, helium is not evenly mixed throughout the ejecta. In addition, a large fraction of this high-velocity helium is  accelerated from its initial location just above the antipodal, core-shell explosion site. Conversely in Case 1, the central detonation, even when off-center, causes a symmetrical mixing of the ejecta along all lines of sight, distributing both helium and nickel among all layers of the ejecta.

\section{Conclusions}\label{conclusions}
Examining the dynamic evolution of the double-detonation scenario subject to increasing spatial resolution highlights a divergence in outcomes with respect to the particular mode of explosion. In both cases, a generally credible thermonuclear supernova occurs, but the paths to explosion are different at a qualitative level and \replaced{lead to several quantitative differences in observable consequences.}{lead to significant quantitative differences in nucleosynthetic yields and the distribution of those yields.}
At coarser resolutions, burning proceeds in a continuous fashion with no particular separation in burning regimes other than the type of fuel being burned. Shell burning proceeds around the limb of the white dwarf without causing the core to ignite until burning reaches the antipodal point of ignition, forming a pinch at the limb of the core which begins carbon burning towards the center.
This edge ignition of the core overwhelms the inward compression wave caused by shell burning, blurring the difference with direct edge-lit cases.
However, at finer resolutions --2~km and better-- a new picture emerges: shell burning proceeds in the same qualitative manner as coarser cases, but it reaches the antipodal point with less heated mass and therefore avoids an antipodal edge ignition.
This causes shell burning to subside, and pressure waves set in motion earlier by shell burning move through the core, prompting a new stage where there is little nuclear burning occurring throughout the domain for approximately 300 milliseconds. Once the pressure waves converge off-center in the core, burning ignites a secondary detonation within it, away from shell burning remnants.\added{The combined effect of a distinct core detonation in both time and space yields clear differences in nucleosynthetic yields from the burning of the core and the symmetry of the ejecta which should be borne out in observations, specifically for line-of-sight arguments and intermediate mass element yields.}
Though burning length scales involved in the physical events are still out of reach for the stated resolutions, the emergence of three clear stages at resolutions under 4 kilometers warrants a minimal resolution for consistency of the general scenario. This notion can be extended to full 3d simulations given that the three stages, though connected, are sufficiently independent from each other to conflate symmetry axis effects or nuclear network picked (energy deposition dependence).
\acknowledgments
This research was supported by the Exascale Computing Project (17-SC-20-SC), a collaborative effort of the U.S. Department of Energy Office of Science and the National Nuclear Security Administration. Research at Oak Ridge National Laboratory is supported under contract DE-AC05-00OR22725 from the U.S. Department of Energy to UT-Battelle, LLC. This research used resources of the Oak Ridge Leadership Computing Facility, which is a DOE Office of Science User Facility supported under contract DE-AC05-00OR22725. This research was supported by the Exascale Computing Project (17-SC-20-SC), a collaborative effort of the U.S. Department of Energy Office of Science and the National Nuclear Security Administration.
\facilities{Summit (\href{https://www.olcf.ornl.gov/}{OLCF})}
\software{\href{https://www.astropy.org/}{Astropy} \citep{2013A&A...558A..33A},
    \href{https://yt-project.org/}{YT} \citep{2011ApJS..192....9T},
    \href{https://flash-x.org/}{FLASH-X} \citep{DUBEY2009512}
}

\bibliography{references}{}
\bibliographystyle{aasjournal}
\listofchanges
\end{document}